\documentclass[a4paper,12pt]{article}
\usepackage{amsmath, amssymb, latexsym, amscd, amsthm,amsfonts,amstext}
\usepackage[mathscr]{eucal}
\usepackage{graphicx}
\usepackage{subfig}

\numberwithin{equation}{section}
\input{tcilatex}
\begin{document}

\date{\vspace{-5ex}}
\title{Numerical solution of an ill-posed Cauchy problem for a quasilinear
parabolic equation using a Carleman weight function}
\author{Michael V. Klibanov$^{\ast }$, Nikolaj A. Koshev$^{\circ }$, Jingzhi
Li$^{\bullet }$ \and and Anatoly G. Yagola$^{\circ \circ }$ \and $^{\ast }$%
Department of Mathematics and Statistics \and University of North Carolina
at Charlotte \and Charlotte, NC 28223, USA \and $^{\circ }$Institute of
Computational Mathematics \and University of S\~{a}n Paulo, S\~{a}o Carlos,
SP 13566-590, Brazil \and $^{\bullet }$Southern University of Science and
Technology of China \and Shenzhen, 518005, People's Republic of China \and $%
^{\circ \circ }$Department of Mathematics, Faculty of Physics \and Moscow
State University, Moscow, 119991, Russian Federation \and E-mails:
mklibanv@uncc.edu, nikolay.koshev@gmail.com, \and li.jz@sustc.edu.cn,
yagola@physics.msu.ru}
\date{}
\maketitle

\begin{abstract}
This is the first publication in which an ill-posed Cauchy problem for a
quasilinear PDE is solved numerically by a rigorous method. More precisely,
we solve the side Cauchy problem for a 1-d quasilinear parabolc equation.
The key idea is to minimize a strictly convex cost functional with the
Carleman Weight Function in it. Previous publications about numerical
solutions of ill-posed Cauchy problems were considering only linear
equations.
\end{abstract}

\graphicspath{
{FIGURES/}
 {pics/}}



\section{Introduction}

\label{sec:1}

This is the first publication in which an ill-posed Cauchy problem for a
quasilinear PDE is solved numerically by a rigorous method. This is done for
a 1d quasilinear parabolic equation with the lateral Cauchy data given on
one edge of the interval. Initial condition is unknown. We implement
numerically the idea of the paper \cite{Klib15} of the first author. It was
proposed in \cite{Klib15} to construct globally strictly convex weighted
Tikhonov-like functional with Carleman Weight Functions (CWFs) in them. In
particular, we demonstrate numerically here that the presence of the CWF
significantly improves the solution accuracy even in the case of the linear
PDE $u_{t}=u_{xx}.$

The topic of numerical solutions of ill-posed Cauchy problems for PDEs is
very popular in the field of ill-posed problems. As some examples, we refer
to, e.g. \cite%
{Bar1,B1,B2,D1,D2,E1,E2,HaoLes,Kab,Karch1,Karch2,KT,APNUM,Kozlov,Bar2} and
there are many more publications on this topic. However, all those works
consider only linear PDEs. Even though the paper \cite{Bar1} considers a
quasilinear equation, in fact that equation can be reduced to a linear one
via a change of variables. Two natural questions therefore are:

\begin{enumerate}
\item Is it possible to develop a numerical method for ill-posed Cauchy
problems for nonlinear PDEs?

\item Can the method of item 1 deliver at least one point in a sufficiently
small neighborhood of the exact solution, provided that no information about
this neighborhood would be given in advance?
\end{enumerate}

These two questions were addressed positively in the paper \cite{Klib15} of
the first author. This was done for those quasilinear PDEs of the second
order, whose principal parts of operators are linear and admit Carleman
estimates. In other words, those are parabolic, elliptic and hyperbolic
quasilinear PDEs with linear principal parts of their operators. However,
numerical experiments were not a part of \cite{Klib15}. So, the current
paper complements \cite{Klib15} in this sense.

Similar ill-posed Cauchy problems for linear parabolic PDEs were considered
in, e.g. \cite{B1,D2,E1,E2,KT}. Ideas, similar to the one of this paper,
were applied in works of the first author with coauthors \cite%
{BK2,Klib97,KNT,KK}. In these works globally strictly convex cost
functionals for Coefficient Inverse Problems (CIPs) were constructed.
Furthermore, the publication \cite{KNT} contains numerical results for the
1-d case.

For the first time, the method of Carleman estimates was introduced in the
field of inverse problems in the paper of Bukhgeim and Klibanov \cite{BukhK}
in 1981. The goal of the publication \cite{BukhK} was to apply Carleman
estimates for proofs of uniqueness and stability results for CIPs. The idea
of \cite{BukhK} became quite popular since then with many publications of a
number of authors. Since this is not a survey of that method, we refer here
to only a limited number of publications \cite%
{BK1,Im1,Im2,IsMilan,Is,KT,Klib,Trig,Yam}. In particular, papers \cite{Klib,
Yam} are surveys.

In section 2 we formulate the problem, describe our numerical method and
also formulate some relevant theorems. In sections 3 and 3 we prove Theorems
1 and 2 respectively. In section 5 we describe our numerical implementation
and in section 6 we present our numerical results.

\section{Statement of the Problem and the Numerical Method}

\label{sec:2}

A general statement of the ill-posed Cauchy problem considered here can be
found in \cite{Klib15}.\ The same about some theorems below, which can be
formulated in more general forms. However, since we consider only the 1-d
case here, we formulate our problems and results for this case only: for
brevity.

\subsection{Statement of the problem}

\label{sec:2.1}

Let $T=const.\in \left( 0,1\right) .$ Denote $Q_{T}^{\pm }=\left( 0,1\right)
\times \left( -T,T\right) .$ Let the function $c\left( x,t\right) \in
C^{1}\left( \overline{Q_{T}^{\pm }}\right) $ and $c_{0}\leq c\left(
x,t\right) \leq c_{1},\forall \left( x,t\right) \in Q_{T}^{\pm },$ where the
numbers $c_{0},c_{1}>0.$ Let the function $P\in C^{1}\left( \mathbb{R}%
^{2}\times \overline{Q_{T}^{\pm }}\right) .$ Consider the following forward
problem in $Q_{T}^{\pm }$ 
\begin{equation}
c\left( x,t\right) u_{t}=u_{xx}+P\left( u_{x},u,x,t\right) \text{ in }%
Q_{T}^{\pm },  \label{2.1}
\end{equation}%
\begin{equation}
u\left( x,-T\right) =f\left( x\right) ,  \label{2.2}
\end{equation}%
\begin{equation}
u\left( 0,t\right) =g\left( t\right) ,u\left( 1,t\right) =p\left( t\right) .
\label{2.3}
\end{equation}%
Uniqueness and existence theorems for this problem are well known, see, e.g.
the book of Ladyzhenskaya, Solonnikov and Ural'ceva \cite{LSU}. So, we
assume that there exists unique solution $u\in C^{2,1}\left( \overline{%
Q_{T}^{\pm }}\right) $ of the problem (\ref{2.1})-(\ref{2.3}). Our interest
is in the following ill-posed Cauchy problem:

\textbf{Ill-Posed Cauchy Problem 1}. \emph{Suppose that functions }$f\left(
x\right) $\emph{\ and }$g\left( t\right) $\emph{\ are unknown whereas the
function }$p\left( t\right) $ \emph{is known. Also, assume that the
following function }$q\left( t\right) $\emph{\ is known}%
\begin{equation}
u_{x}\left( 1,t\right) =q\left( t\right) ,t\in \left( -T,T\right) .
\label{2.4}
\end{equation}%
\emph{Determine the function }$u\left( x,t\right) $\emph{\ in at least a
subdomain of the time cylinder }$Q_{T}^{\pm }.$

Uniqueness of the solution of this problem follows immediately from the well
known uniqueness theorem for a general parabolic PDE of the second order
with the lateral Cauchy data, see, e.g. Chapter 4 of \cite{LRS}.

\subsection{Numerical method}

\label{sec:2.2}

Following \cite{Klib15}, we introduce the Carleman estimates first. This
estimate is different from the one of \cite{Klib15}, since the CWF depends
on two large parameters, instead of just one here. As a result, that CWF
decays too rapidly. So, we have discovered in our computations that the rate
of decay of that CWF is inconvenient for the numerical implementation. Let $%
\lambda >0$ be a large positive parameter. Consider functions $\psi \left(
x,t\right) $ and $\varphi _{\lambda }\left( x,t\right) $ defined as%
\begin{equation}
\varphi _{\lambda }\left( x,t\right) =\exp \left[ \lambda \left(
x^{2}-t^{2}\right) \right] .  \label{2.5}
\end{equation}%
For any $\theta \in \left( 0,1\right) $ let $Q_{T,\theta }^{\pm }=Q_{T}^{\pm
}\cap \left\{ x\in \left( \theta ,1\right) \right\} .$

\textbf{Theorem 1}. \emph{For any }$\theta \in \left( 0,1\right) ,T>0$\emph{%
\ there exists a sufficiently large number }$\lambda _{0}=\lambda _{0}\left(
\theta ,T\right) >1$\emph{\ such that for all }$\lambda \geq \lambda _{0}$%
\emph{\ and for any function }$v\in C^{2,1}\left( \overline{Q_{T,\theta
}^{\pm }}\right) $\emph{\ the following pointwise Carleman estimate is valid 
}%
\begin{equation}
\left( c\left( x,t\right) v_{t}-v_{xx}\right) ^{2}\varphi _{\lambda
}^{2}\geq C\lambda v_{x}^{2}\varphi _{\lambda }^{2}+C\lambda
^{3}v^{2}\varphi _{\lambda }^{2}+U_{x}+V_{t},\forall \left( x,t\right) \in
Q_{T,\theta }^{\pm },  \label{2.6}
\end{equation}%
\emph{where the constant }$C=C\left( Q_{T,\theta }^{\pm },c_{0},\left\Vert
c\right\Vert _{C^{1}\left( \overline{Q_{T,\theta }^{\pm }}\right) },\theta
\right) >0$\emph{\ depends only on listed parameters and is independent on
the function }$v$\emph{. The functions }$U$\emph{\ and }$V$\emph{\ can be
estimated as}%
\begin{equation}
\left\vert U\right\vert ,\left\vert V\right\vert \leq C\lambda ^{3}\left(
v_{x}^{2}+v_{t}^{2}+v^{2}\right) \varphi _{\lambda }^{2}.  \label{2.7}
\end{equation}

For any number $\alpha \in \left( 0,1-T^{2}\right) $ denote 
\begin{equation}
G_{\alpha }=\left\{ \left( x,t\right) \in Q_{T}^{\pm }:x^{2}-t^{2}>\alpha
\right\} .  \label{2.8}
\end{equation}%
Hence, $G_{\alpha }\cap \left\{ t=\pm T\right\} =\varnothing $ and $%
G_{\alpha }\subset Q_{T,\sqrt{\alpha }}^{\pm }.$ The boundary of the domain $%
G_{\alpha }$ is formed by the straight line $\left\{ x=1\right\} $ and the
level curve $\left\{ x^{2}-t^{2}=\alpha \right\} $ of the function $\psi
\left( x,t\right) ,$%
\begin{equation}
\partial G_{\alpha }=\partial _{1}G_{\alpha }\cup \partial _{2}G_{\alpha },
\label{2.9}
\end{equation}%
\begin{equation}
\partial _{1}G_{\alpha }=\left\{ \left( x,t\right) :x=1,\left\vert
t\right\vert <\sqrt{1-\alpha }\right\} ,  \label{2.10}
\end{equation}%
\begin{equation}
\partial _{2}G_{\alpha }=\left\{ \left( x,t\right) :x\in \left( 0,1\right)
,x^{2}-t^{2}=\alpha \right\} .  \label{2.11}
\end{equation}%
Define the operator $A$ and its principal part $A_{0}$ as%
\begin{equation}
A\left( u\right) =c\left( x,t\right) u_{t}-u_{xx}-P\left( u_{x},u,x,t\right)
,\text{ }A_{0}u=c\left( x,t\right) u_{t}-u_{xx}.  \label{200}
\end{equation}%
Fix an $\alpha _{0}\in \left( 0,1-T^{2}\right) $ and let the number $%
\varepsilon >0$ be so small that $\alpha _{0}+2\varepsilon <1-T^{2}.$ Let $%
R>0$ be an arbitrary number. Denote 
\begin{eqnarray}
B\left( R\right) &=&\left\{ u\in H^{3}\left( Q_{T}^{\pm }\right) :\left\Vert
u\right\Vert _{H^{3}\left( Q_{T}^{\pm }\right) }<R,u\left( 1,t\right)
=p\left( t\right) ,u_{x}\left( 1,t\right) =q\left( t\right) \right\} ,
\label{2.110} \\
H_{0}^{3}\left( Q_{T}^{\pm }\right) &=&\left\{ u\in H^{3}\left( Q_{T}^{\pm
}\right) ,u\left( 1,t\right) =0,u_{x}\left( 1,t\right) =0\right\} .
\label{2.111}
\end{eqnarray}%
Here and below all functions are real valued ones. Note that by the
embedding theorem $H^{3}\left( Q_{T}^{\pm }\right) \subset C^{1}\left( 
\overline{Q_{T}^{\pm }}\right) $ and 
\begin{equation}
\left\Vert f\right\Vert _{C^{1}\left( \overline{Q_{T}^{\pm }}\right) }\leq
C\left\Vert f\right\Vert _{H^{3}\left( Q_{T}^{\pm }\right) },\forall f\in
H^{3}\left( Q_{T}^{\pm }\right) .  \label{2.12}
\end{equation}%
Here and below $C=C\left( Q_{T}^{\pm }\right) >0$ denotes different
constants depending only on the domain $Q_{T}^{\pm }$.

Our numerical method consists in the minimization of the weighted functional 
$J_{\lambda ,\beta }$ with the regularization parameter $\beta \in \left(
0,1\right) $ on the set $B\left( R\right) ,$ where%
\begin{equation}
J_{\lambda ,\beta }\left( u\right) =e^{-2\lambda \left( \alpha
_{0}+\varepsilon \right) }\dint\limits_{Q_{T}^{\pm }}\left[ A\left( u\right) %
\right] ^{2}\varphi _{\lambda }^{2}dxdt+\beta \left\Vert u\right\Vert
_{H^{3}\left( Q_{T}^{\pm }\right) }^{2}.  \label{2.13}
\end{equation}

\subsection{Theorems\ \ }

\label{sec:2.3}\ \ \ \ \ 

In principle the convergence of the gradient method is known in the case
when its starting point is located in a small neighborhood of the minimizer.
However, the main point of Theorem 3 is that due to the strict convexity of
the functional $J_{\lambda ,\beta }\left( u\right) $ on the set $B\left(
R\right) ,$ the sequence (\ref{2.16}) converges to the unique minimizer
starting from an arbitrary point $u_{1}\in B\left( R\right) $. Since no
restrictions are imposed on the number $R$, then this is the global
convergence. \ \ \ \ \ \ \ \ \ \ \ \ \ \ \ \ \ \ \ \ \ \ \ \ \ \ \ \ \ \ \ \
\ \ \ \ \ \ \ \ \ \ \ \ \ \ \ \ \ \ \ \ \ \ \ \ \ \ \ \ \ \ \ \ \ \ \ \ \ \
\ \ \ \ \ \ \ \ \ \ \ \ \ \ \ \ \ \ \ \ \ \ \ \ \ \ \ \ \ \ \ \ \ \ \ \ \ \
\ \ \ \ \ \ \ \ \ \ \ \ \ \ \ \ \ \ \ \ \ \ \ \ \ \ \ \ \ \ \ \ \ \ \ \ \ \
\ \ \ \ \ \ \ \ \ \ \ \ \ \ \ \ \ \ \ \ \ \ \ \ \ \ \ \ \ \ \ \ \ \ \ \ \ \
\ \ \ \ \ \ \ \ \ \ \ \ \ \ \ \ \ \ \ \ \ \ \ \ \ \ \ \ \ \ \ \ \ \ \ \ \ \
\ \ \ \ \ \ \ \ \ \ \ \ \ \ \ \ \ \ \ \ \ \ \ \ \ \ \ \ \ \ \ \ \ \ \ \ \ \
\ \ \ \ \ \ \ \ \ \ \ \ \ \ \ \ \ \ \ \ \ \ \ \ \ \ \ \ \ \ \ \ \ \ \ \ \ \
\ \ \ \ \ \ 

We now reformulate theorems 2.1-2.3 of \cite{Klib15} for our case. Since the
function $P\in C^{2}\left( \mathbb{R}^{2}\times \overline{Q_{T}^{\pm }}%
\right) ,$ then for each $R>0$ there exists a constant $M=M\left( R,P\right)
>0$ depending only on the number $R$ and the function $P$ such that 
\begin{equation}
\left\vert \partial _{u_{x}}^{k}\partial _{u}^{s}P\left( u_{x},u,x,t\right)
\right\vert \leq M\left( R,P\right) ,1\leq k+s\leq 2,\forall u\in B\left(
R\right) ,\forall \left( x,t\right) \in Q_{T}^{\pm }\text{ .}  \label{2.14}
\end{equation}

\textbf{Theorem 2}. \emph{Let }$R>0$\emph{\ be an arbitrary number. Then for
every function }$u\in B\left( R\right) $\emph{\ there exists the Fr\'{e}chet
derivative }$J_{\lambda ,\beta }^{\prime }\left( u\right) \in
H_{0}^{3}\left( Q_{T}^{\pm }\right) $\emph{\ of the functional (\ref{2.13}).
Let }$\lambda _{0}=\lambda _{0}\left( \sqrt{\alpha _{0}},T\right) >1$\emph{\
be the parameter of Theorem 1. Then there exists a sufficiently large number 
}$\lambda _{1}=\lambda _{1}\left( \alpha _{0},M,R,T,\varepsilon \right) \geq
\lambda _{0}$\emph{\ such that for all }$\lambda \geq \lambda _{1}$\emph{\
and for every }$\beta \in \left( e^{-\lambda \varepsilon },1\right) $\emph{\
the functional }$J_{\lambda ,\beta }$\emph{\ is strictly convex on the set }$%
B\left( R\right) ,$\emph{\ i.e.}%
\begin{equation}
J_{\lambda ,\beta }\left( u_{2}\right) -J_{\lambda ,\beta }\left(
u_{1}\right) -J_{\lambda ,\beta }^{\prime }\left( u_{1}\right) \left(
u_{2}-u_{1}\right)  \label{2.15}
\end{equation}%
\begin{equation*}
\geq C_{1}e^{2\lambda \varepsilon }\left\Vert u_{2}-u_{1}\right\Vert
_{H^{1,0}\left( G_{\alpha _{0}+2\varepsilon }\right) }^{2}+\frac{\beta }{2}%
\left\Vert u_{2}-u_{1}\right\Vert _{H^{3}\left( Q_{T}^{\pm }\right) }^{2},
\end{equation*}%
\emph{where the constant }$C_{1}=C_{1}\left( \alpha _{0},M,R\right) >0$\emph{%
\ depends only on listed parameters.}

Here the space $H^{1,0}\left( G_{\alpha _{0}+2\varepsilon }\right) $ is the
Hilbert space of real valued functions with the norm%
\begin{equation*}
\left\Vert u\right\Vert _{H^{1,0}\left( G_{\alpha _{0}+2\varepsilon }\right)
}=\sqrt{\left\Vert u\right\Vert _{L_{2}\left( G_{\alpha _{0}+2\varepsilon
}\right) }^{2}+\left\Vert u_{x}\right\Vert _{L_{2}\left( G_{\alpha
_{0}+2\varepsilon }\right) }^{2}}.
\end{equation*}%
We now construct the gradient method of the minimization of the functional (%
\ref{2.13}) on the set $B\left( R\right) $. For brevity we consider only the
simplest version of that method. Consider an arbitrary point $u_{1}\in
B\left( R\right) $. Let $\gamma >0$ be the step size of the gradient method.
Then the sequence $\left\{ u_{n}\right\} _{n=1}^{\infty }$ of the gradient
method is%
\begin{equation}
u_{n+1}=u_{n}-\gamma J_{\lambda ,\beta }^{\prime }\left( u_{n}\right)
,n=1,2,...  \label{2.16}
\end{equation}%
For brevity, we do not indicate here and in some places below dependencies
of some functions on the parameter $\lambda .$ Theorem 3 claims the
convergence of the sequence (\ref{2.16}).

\textbf{Theorem 3}. \emph{Let conditions of Theorem 2 hold.} \emph{Let }$%
\lambda _{1}$\emph{\ be the parameter of Theorem 2. Let }$\lambda \geq
\lambda _{1}$\emph{\ and }$\beta \in \left( e^{-\lambda \varepsilon
},1\right) .$ \emph{Assume that the functional }$J_{\lambda ,\beta }$\emph{\
achieves its minimal value on the set }$B\left( R\right) $\emph{\ at a point 
}$u_{\min }\in B\left( R\right) ,$\emph{\ which we call \textquotedblleft
minimizer". Then the minimizer is unique on }$B\left( R\right) .$\emph{\
Assume that the sequence }$\left\{ u_{n}\right\} _{n=1}^{\infty }\subset
B\left( R\right) ,$\emph{\ where }$u_{1}$\emph{\ is an arbitrary point of }$%
B\left( R\right) .$\emph{\ Then there exists a sufficiently small number }$%
\gamma =\gamma \left( \lambda ,\beta ,\alpha _{0},M,R,T\right) \in \left(
0,1\right) $\emph{\ depending only on listed parameters and \ a number }$%
r=r\left( \gamma \right) \in \left( 0,1\right) $\emph{\ such that the
sequence }$\left\{ u_{n}\right\} _{n=1}^{\infty }$\emph{\ converges to the
point }$u_{\min }$\emph{\ in the norm of the space }$H^{3}\left( Q_{T}^{\pm
}\right) $\emph{\ and the following convergence estimate holds}%
\begin{equation*}
\left\Vert u_{n+1}-u_{\min }\right\Vert _{H^{3}\left( Q_{T}^{\pm }\right)
}\leq r^{n}\left\Vert u_{1}-u_{\min }\right\Vert _{H^{3}\left( Q_{T}^{\pm
}\right) },n=1,2,...
\end{equation*}

The minimizer $u_{\min }\in B\left( R\right) $ is called the
\textquotedblleft regularized solution" in the regularization theory \cite%
{BK1,T}. The next natural question is about the convergence of regularized
solutions to the exact solution. We now modify the material of pages 6,7 of 
\cite{Klib15}, where this question was addressed for a general case. In
accordance with the Tikhonov concept for ill-posed problems \cite{BK1,T}, we
assume that there exists an exact solution $u^{\ast }\in H^{3}\left(
Q_{T}^{\pm }\right) $ of our problem with noiseless data $p^{\ast }\left(
t\right) $ and $q^{\ast }\left( t\right) .$ In other words, we assume the
there exists the solution $u^{\ast }\left( x,t\right) $ of the following
problem%
\begin{equation}
c\left( x,t\right) u_{t}^{\ast }=u_{xx}^{\ast }+P\left( u_{x}^{\ast
},u^{\ast },x,t\right) \text{ in }Q_{T}^{\pm },  \label{2.17}
\end{equation}%
\begin{equation}
u^{\ast }\left( 1,t\right) =p^{\ast }\left( t\right) ,u_{x}^{\ast }\left(
1,t\right) =q^{\ast }\left( t\right) ,t\in \left( -T,T\right) ,  \label{2.18}
\end{equation}%
\begin{equation}
u^{\ast }\in H^{3}\left( Q_{T}^{\pm }\right) ,\text{ }p^{\ast },q^{\ast }\in
H^{3}\left( -T,T\right) .  \label{2.19}
\end{equation}%
As to the functions $p$ and $q$ in (\ref{2.3}) and (\ref{2.4}), we assume $%
p,q\in H^{3}\left( -T,T\right) $ and that they are given with an error of
the level $\delta ,$ i.e.%
\begin{equation}
\left\Vert p-p^{\ast }\right\Vert _{H^{3}\left( -T,T\right) }\leq \delta
,\left\Vert q-q^{\ast }\right\Vert _{H^{3}\left( -T,T\right) }\leq \delta .
\label{2.20}
\end{equation}%
Next, following \cite{Klib15}, we construct functions $F,F^{\ast }\in
H^{3}\left( Q_{T}^{\pm }\right) $ as%
\begin{equation}
F\left( x,t\right) =p\left( t\right) +\left( x-1\right) q\left( t\right)
,F^{\ast }\left( x,t\right) =p^{\ast }\left( t\right) +\left( x-1\right)
q^{\ast }\left( t\right) .  \label{2.21}
\end{equation}%
Hence, 
\begin{equation*}
F\left( 1,t\right) =p\left( t\right) ,F_{x}\left( 1,t\right) =q\left(
t\right) ,F^{\ast }\left( 1,t\right) =p^{\ast }\left( t\right) ,F_{x}^{\ast
}\left( 1,t\right) =q^{\ast }\left( t\right) ,
\end{equation*}%
as it is required in \cite{Klib15}. Furthermore, (\ref{2.20}) and (\ref{2.21}%
) imply the following analog of the estimate (2.29) in \cite{Klib15} is
valid 
\begin{equation*}
\left\Vert F-F^{\ast }\right\Vert _{H^{3}\left( Q_{T}^{\pm }\right) }\leq
C\delta .
\end{equation*}%
We now can formulate Theorem 4 about the convergence of regularized
solutions. This theorem is a direct analog of Theorem 2.3 of \cite{Klib15}.
To be in the agreement with (2.13) of \cite{Klib15}, we note that $\max_{%
\overline{Q}_{T}^{\pm }}\left( x^{2}-t^{2}\right) =1.$

\textbf{Theorem 4}. \emph{Let conditions of Theorems 2, 3 hold. Let the
function }$u^{\ast }$\emph{\ be the solution of the problem (\ref{2.17})-(%
\ref{2.19}). Assume that inequalities (\ref{2.20}) are valid. Let the
parameter }$\lambda _{1}$\emph{\ be the same as in Theorem 2. Then there
exists a number }$\lambda _{2}=\lambda _{2}\left( \alpha
_{0},M,R,T,\varepsilon \right) \geq \lambda _{1}$\emph{\ and the number }$%
C_{2}=C_{2}\left( \alpha _{0},M,R,T,\varepsilon \right) >0$\emph{, both
depending only on listed parameters, such that if the number }$\delta
_{0}\in \left( 0,e^{-4\lambda _{2}}\right) ,$\emph{\ then for all }$\lambda
\geq \lambda _{2},\delta \in \left( 0,\delta _{0}\right) ,\beta \in \left(
e^{-\lambda \varepsilon },1\right) $\emph{\ the following estimates are valid%
}%
\begin{equation}
\left\Vert u^{\ast }-u_{\min }\right\Vert _{H^{1,0}\left( G_{\alpha
_{0}+2\varepsilon }\right) }\leq C_{3}\delta ^{\varepsilon /4},  \label{2.22}
\end{equation}%
\begin{equation}
\left\Vert u_{n+1}-u^{\ast }\right\Vert _{H^{1,0}\left( G_{\alpha
_{0}+2\varepsilon }\right) }\leq C_{3}\delta ^{\varepsilon
/4}+r^{n}\left\Vert u_{1}-u_{\min }\right\Vert _{H^{3}\left( Q_{T}^{\pm
}\right) }.  \label{2.23}
\end{equation}

Even though the convergence here is in a subdomain of the domain $Q_{T}^{\pm
},$ this seems to be sufficient for computations. The combination of
Theorems 2,3,4 addresses two questions posed in the beginning of section 1.
Now about proofs of above theorems. As to Theorem 1, it is known from the
survey of Yamamoto \cite{Yam}. However, since a general parabolic operator
of the second order is considered in \cite{Yam}, we prove this theorem below
for our specific operator $c\left( x,t\right) \partial _{t}-\partial _{x}^{2}
$: for the sake of completeness. As to Theorem 2, it is a direct analog of
theorem 2.1 of \cite{Klib15}. However, there is an important difference too.
The domain of integration in \cite{Klib15} is $G_{\alpha }.$ On the other
hand, it is more convenient for computations to integrate over the entire
rectangle $Q_{T}^{\pm }$ as in (\ref{2.12}). This means that the we need to
prove Theorem 2. We do not prove Theorem 3 here, since its direct analog was
proved in \cite{BK2}. Also, we do not prove Theorem 4 below, since its
direct analogs were proved in \cite{BK2} and \cite{Klib15}.

\section{Proof of Theorem 1}

\label{sec:3}

Denote $w\left( x,t\right) =v\left( x,t\right) \exp \left[ \lambda \left(
x^{2}-t^{2}\right) \right] .$ Then $v=w\exp \left[ -\lambda \left(
x^{2}-t^{2}\right) \right] .$ Express derivatives of $v$ via derivatives of $%
w$. We obtain%
\begin{equation*}
v_{t}=\left( w_{t}+2\lambda tw\right) \exp \left[ -\lambda \left(
x^{2}-t^{2}\right) \right] ,\text{ }v_{x}=\left( w_{x}-2\lambda xw\right)
\exp \left[ -\lambda \left( x^{2}-t^{2}\right) \right] ,
\end{equation*}%
\begin{equation*}
v_{xx}=\left( w_{xx}-4\lambda xw_{x}+4\lambda ^{2}x^{2}w\right) \exp \left[
-\lambda \left( x^{2}-t^{2}\right) \right] .
\end{equation*}%
Hence,%
\begin{equation*}
\left( A_{0}v\right) ^{2}\varphi _{\lambda }^{2}=\left( v_{t}-v_{xx}\right)
^{2}\varphi _{\lambda }^{2}=\left[ \left( w_{t}+4\lambda xw_{x}\right)
-\left( w_{xx}+4\lambda ^{2}x^{2}w-2\lambda tw\right) \right] ^{2}
\end{equation*}%
\begin{equation}
\geq \left( -2w_{t}-8\lambda xw_{x}\right) \left( w_{xx}+4\lambda
^{2}x^{2}w-2\lambda tw\right) .  \label{3.1}
\end{equation}%
First, we work with the term $-2w_{t}\left( w_{xx}+4\lambda
^{2}x^{2}w\right) $ in (\ref{3.1}). We obtain%
\begin{equation*}
-2w_{t}\left( w_{xx}+4\lambda ^{2}x^{2}w-2\lambda tw\right) =\left(
-2w_{t}w_{x}\right) _{x}+2w_{tx}w_{x}+\left( -4\lambda
^{3}x^{2}w^{2}+2\lambda tw^{2}\right) _{t}-2\lambda w^{2}
\end{equation*}%
\begin{equation}
=\left( -2w_{t}w_{x}\right) _{x}+\left( w_{x}^{2}-4\lambda
^{3}x^{2}w^{2}+2\lambda tw^{2}\right) _{t}-2\lambda w^{2}.  \label{3.2}
\end{equation}%
Next, we work with the term $-8\lambda xw_{x}\left( w_{xx}+4\lambda
^{2}x^{2}w-2\lambda tw\right) $ in (\ref{3.1}). We obtain%
\begin{equation*}
-8\lambda xw_{x}\left( w_{xx}+4\lambda ^{2}x^{2}w-2\lambda tw\right) =\left(
-4\lambda xw_{x}^{2}\right) _{x}+4\lambda w_{x}^{2}
\end{equation*}%
\begin{equation}
+\left( -16\lambda ^{3}x^{3}w^{2}+8\lambda ^{2}xtw^{2}\right) _{x}+48\lambda
^{3}x^{2}w^{2}-8\lambda ^{2}tw^{2}.  \label{3.3}
\end{equation}%
Choose the parameter $\lambda _{0}=\lambda _{0}\left( \theta ,T\right) >1$
so large that $24\lambda ^{3}\theta ^{2}>8\lambda ^{2}T+2\lambda ,\forall
\lambda \geq \lambda _{0}.$ Then summing up (\ref{3.2}) and (\ref{3.3}) and
taking into account (\ref{3.1}), we obtain for these values of $\lambda $%
\begin{equation*}
\left( v_{t}-v_{xx}\right) ^{2}\varphi _{\lambda }^{2}\geq 4\lambda
w_{x}^{2}+24\lambda ^{3}\theta ^{2}w^{2}
\end{equation*}%
\begin{equation*}
+\left( -2w_{t}w_{x}-4\lambda xw_{x}^{2}-16\lambda ^{3}x^{3}w^{2}+8\lambda
^{2}xtw^{2}\right) _{x}+\left( w_{x}^{2}-4\lambda ^{3}x^{2}w^{2}+2\lambda
tw^{2}\right) _{t}.
\end{equation*}%
Next, replacing here $w$ with $v=w\exp \left[ -\lambda \left(
x^{2}-t^{2}\right) \right] ,$ we easily obtain the desired estimates (\ref%
{2.6}) and (\ref{2.7}). $\square $

\section{Proof of Theorem 2}

\label{sec:4}

In this proof $C_{1}=C_{1}\left( \alpha _{0},M,R\right) >0$ denotes
different constants depending only on listed parameters. First, recall that
for all appropriate functions $f\left( y\right) $ of one variable $y\in 
\mathbb{R}$ the following Lagrange formula is valid%
\begin{equation}
f\left( y+z\right) =f\left( y\right) +f^{\prime }\left( y\right) z+\frac{%
z^{2}}{2}f^{\prime \prime }\left( \xi \right) ,\forall y,z\in \mathbb{R},
\label{4.1}
\end{equation}%
where the number $\xi $ is located between numbers $y$ and $y+z$. Let $%
u_{1},u_{2}\in B\left( R\right) $ be two arbitrary functions. Let $%
h=u_{2}-u_{1}.$ Then $u_{2}=u_{1}+h$ and (\ref{2.110})-(\ref{2.12}) imply
that%
\begin{equation}
h\in H_{0}^{3}\left( Q_{T}^{\pm }\right) ,\left\Vert h\right\Vert
_{H^{3}\left( Q_{T}^{\pm }\right) }<2R,\left\Vert h\right\Vert _{C^{1}\left( 
\overline{Q_{T}^{\pm }}\right) }\leq CR.  \label{4.2}
\end{equation}%
Consider the expression for $A\left( u_{1}+h\right) =A\left( u_{2}\right) ,$
where the operator $A$ is defined in (\ref{200}). We have 
\begin{equation}
A\left( u_{1}+h\right) =c\left( x,t\right) \left( u_{1}+h\right) _{t}-\left(
u_{1}+h\right) _{xx}-P\left( u_{1x}+h_{x},u_{1}+h,x,t\right)  \label{4.3}
\end{equation}%
\begin{equation*}
=A_{0}\left( u_{1}\right) +A_{0}\left( h\right) -P\left(
u_{1x}+h_{x},u_{1}+h,x,t\right) .
\end{equation*}%
We now work with the term $P\left( u_{1x}+h_{x},u_{1}+h,x,t\right) $ in (\ref%
{4.3}). Using (\ref{2.14}) and (\ref{4.1}), we obtain in a standard manner 
\begin{equation*}
P\left( u_{1x}+h_{x},u_{1}+h,x,t\right) =
\end{equation*}%
\begin{equation}
P\left( u_{1x},u_{1},x,t\right) +h_{x}\partial _{u_{x}}P\left(
u_{1x},u_{1},x,t\right) +h\partial _{u}P\left( u_{1x},u_{1},x,t\right) +%
\widetilde{P}\left( u_{1x},u_{1},h_{x},h,x,t\right) ,  \label{4.4}
\end{equation}%
where $\widetilde{P}$ is a continuous function of its variables for which
the following estimate holds for all functions $u_{1}\in B\left( R\right) $
and for all functions $h$ satisfying (\ref{4.2}) 
\begin{equation}
\left\vert \widetilde{P}\left( u_{1x},u_{1},h_{x},h,x,t\right) \right\vert
\leq C_{1}\left( h_{x}^{2}+h^{2}\right) ,\forall \left( x,t\right) \in
Q_{T}^{\pm }.  \label{4.5}
\end{equation}%
Hence, using (\ref{4.3}) and (\ref{4.4}), we obtain%
\begin{eqnarray*}
A\left( u_{1}+h\right) &=&A\left( u_{1}\right) +\left[ A_{0}\left( h\right)
+\partial _{u_{x}}P\left( u_{1x},u_{1},x,t\right) h_{x}+\partial _{u}P\left(
u_{1x},u_{1},x,t\right) h\right] \\
&&+\widetilde{P}\left( u_{1x},u_{1},h_{x},h,x,t\right) .
\end{eqnarray*}%
Hence,%
\begin{equation*}
\left[ A\left( u_{1}+h\right) \right] ^{2}-\left[ A\left( u_{1}\right) %
\right] ^{2}
\end{equation*}%
\begin{equation*}
=2A\left( u_{1}\right) \left[ A_{0}\left( h\right) +\partial _{u_{x}}P\left(
u_{1x},u_{1},x,t\right) h_{x}+\partial _{u}P\left( u_{1x},u_{1},x,t\right) h%
\right]
\end{equation*}%
\begin{equation}
+\left[ A_{0}\left( h\right) +\partial _{u_{x}}P\left(
u_{1x},u_{1},x,t\right) h_{x}+\partial _{u}P\left( u_{1x},u_{1},x,t\right) h%
\right] ^{2}+\widetilde{P}^{2}  \label{4.6}
\end{equation}%
\begin{equation*}
+2\left[ A\left( u_{1}\right) +A_{0}\left( h\right) +\partial
_{u_{x}}P\left( u_{1x},u_{1},x,t\right) h_{x}+\partial _{u}P\left(
u_{1x},u_{1},x,t\right) h\right] \widetilde{P}.
\end{equation*}%
The expression in the second line of (\ref{4.6}), which we denote as $%
Z\left( u_{1}\right) \left( h\right) ,$ is linear with respect to $h$.

Consider the linear functional $\overline{J}_{\lambda ,\beta }\left(
u_{1}\right) \left( \eta \right) :H_{0}^{3}\left( Q_{T}^{\pm }\right)
\rightarrow \mathbb{R}$ defined as%
\begin{equation}
\overline{J}_{\lambda ,\beta }\left( u_{1}\right) \left( \eta \right)
=\dint\limits_{Q_{T}^{\pm }}Z\left( u_{1}\right) \left( \eta \right) \varphi
_{\lambda }^{2}dxdt+2\beta \left[ u_{1},\eta \right] ,\text{ }\forall \eta
\in H_{0}^{3}\left( Q_{T}^{\pm }\right) ,  \label{4.7}
\end{equation}%
where $\left[ ,\right] $ denotes the scalar product in $H_{0}^{3}\left(
Q_{T}^{\pm }\right) .$ Then it can be proved similarly with \cite{Klib15}
that $\overline{J}_{\lambda ,\beta }\left( u_{1}\right) \left( \eta \right) $
defines the Fr\'{e}chet derivative $J_{\lambda ,\beta }^{\prime }\left(
u_{1}\right) $ of the functional $J_{\lambda ,\beta }$ at the point $u_{1}.$
More precisely, there exists unique function $M\left( u_{1}\right) \in
H_{0}^{3}\left( Q_{T}^{\pm }\right) $ such that 
\begin{eqnarray}
\overline{J}_{\lambda ,\beta }\left( u_{1}\right) \left( \eta \right) &=&%
\left[ M\left( u_{1}\right) ,\eta \right] ,\forall \eta \in H_{0}^{3}\left(
Q_{T}^{\pm }\right) ,  \label{4.8} \\
M\left( u_{1}\right) &=&J_{\lambda ,\beta }^{\prime }\left( u_{1}\right) \in
H_{0}^{3}\left( Q_{T}^{\pm }\right) .  \label{4.9}
\end{eqnarray}

Hence, using (\ref{2.13}), (\ref{4.5})-(\ref{4.9}) and the Cauchy-Schwarz
inequality, we obtain 
\begin{equation*}
J_{\lambda ,\beta }\left( u_{1}+h\right) -J_{\lambda ,\beta }\left(
u_{1}\right) -J_{\lambda ,\beta }^{\prime }\left( u_{1}\right) \left(
h\right)
\end{equation*}%
\begin{equation}
\geq \frac{1}{2}e^{-2\lambda \left( \alpha _{0}+\varepsilon \right)
}\dint\limits_{Q_{T}^{\pm }}\left[ A_{0}\left( h\right) \right] ^{2}\varphi
_{\lambda }^{2}dxdt  \label{4.10}
\end{equation}%
\begin{equation*}
-C_{1}e^{-2\lambda \left( \alpha _{0}+\varepsilon \right)
}\dint\limits_{Q_{T}^{\pm }}\left( h_{x}^{2}+h^{2}\right) \varphi _{\lambda
}^{2}dxdt+\beta \left\Vert h\right\Vert _{H^{3}\left( Q_{T}^{\pm }\right)
}^{2}.
\end{equation*}%
Since $\varphi _{\lambda }^{2}\left( x,t\right) <e^{2\lambda \alpha _{0}}$
for $\left( x,t\right) \in Q_{T}^{\pm }\diagdown G_{\alpha _{0}},$ then%
\begin{equation*}
-C_{1}e^{-2\lambda \left( \alpha _{0}+\varepsilon \right)
}\dint\limits_{Q_{T}^{\pm }}\left( h_{x}^{2}+h^{2}\right) \varphi _{\lambda
}^{2}dxdt=-C_{1}e^{-2\lambda \left( \alpha _{0}+\varepsilon \right)
}\dint\limits_{G_{\alpha _{0}}}\left( h_{x}^{2}+h^{2}\right) \varphi
_{\lambda }^{2}dxdt
\end{equation*}%
\begin{equation}
--C_{1}e^{-2\lambda \left( \alpha _{0}+\varepsilon \right)
}\dint\limits_{Q_{T}^{\pm }\diagdown G_{\alpha _{0}}}\left(
h_{x}^{2}+h^{2}\right) \varphi _{\lambda }^{2}dxdt  \label{4.11}
\end{equation}%
\begin{equation*}
\geq -C_{1}e^{-2\lambda \left( \alpha _{0}+\varepsilon \right)
}\dint\limits_{G_{\alpha _{0}}}\left( h_{x}^{2}+h^{2}\right) \varphi
_{\lambda }^{2}dxdt-C_{1}e^{-2\lambda \varepsilon }\left\Vert h\right\Vert
_{H^{1,0}\left( Q_{T}^{\pm }\right) }^{2}.
\end{equation*}%
Next, since $G_{\alpha _{0}}\subset Q_{T}^{\pm },$ then 
\begin{equation}
\frac{1}{2}e^{-2\lambda \left( \alpha _{0}+\varepsilon \right)
}\dint\limits_{Q_{T}^{\pm }}\left[ A_{0}\left( h\right) \right] ^{2}\varphi
_{\lambda }^{2}dxdt\geq \frac{1}{2}e^{-2\lambda \left( \alpha
_{0}+\varepsilon \right) }\dint\limits_{G_{\alpha _{0}}}\left[ A_{0}\left(
h\right) \right] ^{2}\varphi _{\lambda }^{2}dxdt.  \label{4.12}
\end{equation}%
Combining (\ref{4.10})-(\ref{4.12}) and using $\left\Vert h\right\Vert
_{H^{1,0}\left( Q_{T}^{\pm }\right) }^{2}\leq C_{1}\left\Vert h\right\Vert
_{H^{3}\left( Q_{T}^{\pm }\right) }^{2}$, we obtain%
\begin{equation*}
J_{\lambda ,\beta }\left( u_{1}+h\right) -J_{\lambda ,\beta }\left(
u_{1}\right) -J_{\lambda ,\beta }^{\prime }\left( u_{1}\right) \left(
h\right)
\end{equation*}%
\begin{equation}
\geq \frac{1}{2}e^{-2\lambda \left( \alpha _{0}+\varepsilon \right)
}\dint\limits_{G_{\alpha _{0}}}\left[ A_{0}\left( h\right) \right]
^{2}\varphi _{\lambda }^{2}dxdt-C_{1}e^{-2\lambda \left( \alpha
_{0}+\varepsilon \right) }\dint\limits_{G_{\alpha _{0}}}\left(
h_{x}^{2}+h^{2}\right) \varphi _{\lambda }^{2}dxdt  \label{4.13}
\end{equation}%
\begin{equation*}
-C_{1}e^{-2\lambda \varepsilon }\left\Vert h\right\Vert _{H^{3}\left(
Q_{T}^{\pm }\right) }^{2}+\beta \left\Vert h\right\Vert _{H^{3}\left(
Q_{T}^{\pm }\right) }^{2}.
\end{equation*}

Now we use Theorem 1. Integrating estimate (\ref{2.6}) over $G_{\alpha _{0}}$
and using density arguments, we conclude that we can substitute in those
integrals any function $\widetilde{v}\in H^{3}\left( Q_{T,\theta }^{\pm
}\right) $ instead of $v\in C^{2,1}\left( \overline{Q_{T,\theta }^{\pm }}%
\right) .$ Hence, for all $\lambda \geq \lambda _{0}$ 
\begin{equation*}
\frac{1}{2}e^{-2\lambda \left( \alpha _{0}+\varepsilon \right)
}\dint\limits_{G_{\alpha _{0}}}\left[ A_{0}\left( h\right) \right]
^{2}\varphi _{\lambda }^{2}dxdt
\end{equation*}%
\begin{equation}
\geq Ce^{-2\lambda \left( \alpha _{0}+\varepsilon \right)
}\dint\limits_{G_{\alpha _{0}}}\left( \lambda h_{x}^{2}+\lambda
^{3}h^{2}\right) \varphi _{\lambda }^{2}dxdt-C\lambda ^{3}e^{-2\lambda
\left( \alpha _{0}+\varepsilon \right) }\dint\limits_{\partial _{2}G_{\alpha
_{0}}}\left( h_{x}^{2}+h_{t}^{2}+h^{2}\right) \varphi _{\lambda }^{2}dS.
\label{4.14}
\end{equation}%
Since $\varphi _{\lambda }^{2}\left( x,t\right) =e^{2\lambda \alpha _{0}}$
for $\left( x,t\right) \in \partial _{2}G_{\alpha _{0}},$ then (\ref{4.14})
becomes%
\begin{equation*}
\frac{1}{2}e^{-2\lambda \left( \alpha _{0}+\varepsilon \right)
}\dint\limits_{G_{\alpha _{0}}}\left[ A_{0}\left( h\right) \right]
^{2}\varphi _{\lambda }^{2}dxdt
\end{equation*}%
\begin{equation}
\geq Ce^{-2\lambda \left( \alpha _{0}+\varepsilon \right)
}\dint\limits_{G_{\alpha _{0}}}\left( \lambda h_{x}^{2}+\lambda
^{3}h^{2}\right) \varphi _{\lambda }^{2}dxdt-C\lambda ^{3}e^{-2\lambda
\varepsilon }\dint\limits_{\partial _{2}G_{\alpha _{0}}}\left(
h_{x}^{2}+h_{t}^{2}+h^{2}\right) dS  \label{4.15}
\end{equation}%
\begin{equation*}
\geq Ce^{-2\lambda \left( \alpha _{0}+\varepsilon \right)
}\dint\limits_{G_{\alpha _{0}}}\left( \lambda h_{x}^{2}+\lambda
^{3}h^{2}\right) \varphi _{\lambda }^{2}dxdt-C_{1}\lambda ^{3}e^{-2\lambda
\varepsilon }\left\Vert h\right\Vert _{H^{3}\left( Q_{T}^{\pm }\right) }^{2}.
\end{equation*}%
Combining (\ref{4.13})-(\ref{4.15}), we obtain%
\begin{equation*}
J_{\lambda ,\beta }\left( u_{1}+h\right) -J_{\lambda ,\beta }\left(
u_{1}\right) -J_{\lambda ,\beta }^{\prime }\left( u_{1}\right) \left(
h\right)
\end{equation*}%
\begin{equation}
\geq Ce^{-2\lambda \left( \alpha _{0}+\varepsilon \right)
}\dint\limits_{G_{\alpha _{0}}}\left( \lambda h_{x}^{2}+\lambda
^{3}h^{2}\right) \varphi _{\lambda }^{2}dxdt-C_{1}e^{-2\lambda \left( \alpha
_{0}+\varepsilon \right) }\dint\limits_{G_{\alpha _{0}}}\left(
h_{x}^{2}+h^{2}\right) \varphi _{\lambda }^{2}dxdt  \label{4.16}
\end{equation}%
\begin{equation*}
-C_{1}\lambda ^{3}e^{-2\lambda \varepsilon }\left\Vert h\right\Vert
_{H^{3}\left( Q_{T}^{\pm }\right) }^{2}+\beta \left\Vert h\right\Vert
_{H^{3}\left( Q_{T}^{\pm }\right) }^{2}.
\end{equation*}%
Hence, there exists a sufficiently large number $\lambda _{1}=\lambda
_{1}\left( \alpha _{0},M,R,T,\varepsilon \right) \geq \lambda _{0}$\emph{\ }%
such that for all $\lambda \geq \lambda _{1}$ and for every $\beta \in
\left( e^{-\lambda \varepsilon },1\right) $ the first term in the second
line of (\ref{4.16}) absorbs the second term in this line and also the
second term in the third line of (\ref{4.16}) absorbs the first term in this
line. Hence, 
\begin{equation*}
J_{\lambda ,\beta }\left( u_{1}+h\right) -J_{\lambda ,\beta }\left(
u_{1}\right) -J_{\lambda ,\beta }^{\prime }\left( u_{1}\right) \left(
h\right)
\end{equation*}%
\begin{equation}
\geq C_{1}e^{-2\lambda \left( \alpha _{0}+\varepsilon \right) }\lambda
\dint\limits_{G_{\alpha _{0}}}\left( h_{x}^{2}+h^{2}\right) \varphi
_{\lambda }^{2}dxdt+\frac{\beta }{2}\left\Vert h\right\Vert _{H^{3}\left(
Q_{T}^{\pm }\right) }^{2}.  \label{4.17}
\end{equation}%
Next, since $G_{\alpha _{0}+2\varepsilon }\subset G_{\alpha _{0}}$ and since 
$\varphi _{\lambda }^{2}\left( x,t\right) >e^{2\lambda \left( \alpha
_{0}+2\varepsilon \right) }$ for $\left( x,t\right) \in G_{\alpha
_{0}+2\varepsilon },$ then (\ref{4.17}) implies that 
\begin{equation*}
J_{\lambda ,\beta }\left( u_{1}+h\right) -J_{\lambda ,\beta }\left(
u_{1}\right) -J_{\lambda ,\beta }^{\prime }\left( u_{1}\right) \left(
h\right) \geq C_{1}e^{2\lambda \varepsilon }\left\Vert h\right\Vert
_{H^{1,0}\left( G_{\alpha _{0}+2\varepsilon }\right) }^{2}+\frac{\beta }{2}%
\left\Vert h\right\Vert _{H^{3}\left( Q_{T}^{\pm }\right) }^{2}.\text{ \ \ \
\ \ }\square
\end{equation*}

\section{Numerical Implementation}

\label{sec:5}

\subsection{The forward problem}

\label{sec:5.1}

Recall that $Q_{1/2}^{\pm }=\left\{ \left( x,t\right) :x\in \left(
0,1\right) ,t\in \left( -1/2,1/2\right) \right\} .$ For our numerical
testing we have considered the following forward problem:%
\begin{equation}
u_{t}=u_{xx}+aS\left( u\right) +F\left( x,t\right) ,\left( x,t\right) \in
Q_{1/2}^{\pm },  \label{5.1}
\end{equation}%
\begin{equation}
u\left( x,-1/2\right) =f\left( x\right) ,  \label{5.2}
\end{equation}%
\begin{equation}
u\left( 0,t\right) =g\left( t\right) ,  \label{5.3}
\end{equation}%
\begin{equation}
u\left( 1,t\right) =p\left( t\right) .  \label{5.30}
\end{equation}%
In (\ref{5.1}) the number $a=const.\geq 0$ characterizes the degree of the nonlinearity.
For example, $a=0$ corresponds to the linear case. We have chosen two
functions $S\left( u\right) $ in our numerical tests. Our specific functions
in (\ref{5.1})-(\ref{5.30}) were:%
\begin{equation}
S_{1}\left( u\right) =\sin ^{2}\left( u\right) ,S_{2}\left( u\right) =\exp
\left( 0.4u\right) ,  \label{5.4}
\end{equation}%
\begin{equation}
F\left( x,t\right) =10\sin \left[ 100\left( \left( x-0.5\right)
^{2}+t^{2}\right) \right] ,  \label{5.5}
\end{equation}%
\begin{equation}
f\left( x\right) =10\left( x-x^{2}\right) ,  \label{5.6}
\end{equation}%
\begin{equation}
g\left( t\right) =10\sin \left[ 10\left( t-0.5\right) \left( t+0.5\right) %
\right] ,p\left( t\right) =\sin \left[ 10\left( t+0.5\right) \right] .
\label{5.7}
\end{equation}

\begin{figure}[ht!]
	\begin{center}
		\begin{tabular}{ccc}
			\includegraphics[scale=0.45,clip=]{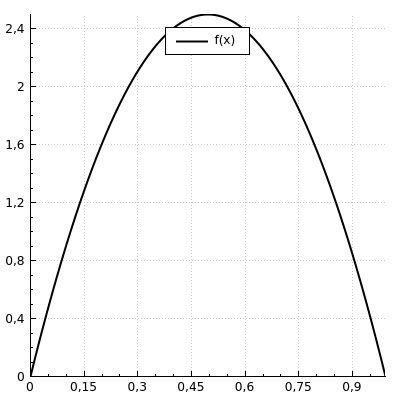} &
			\includegraphics[scale=0.45,clip=]{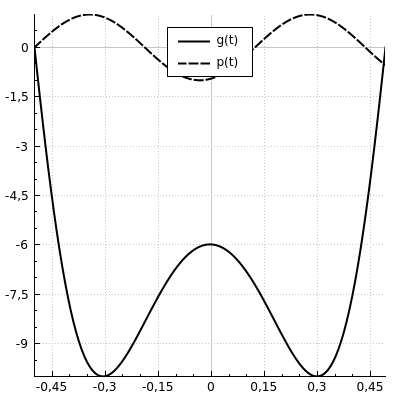} &
			\includegraphics[scale=0.45,clip=]{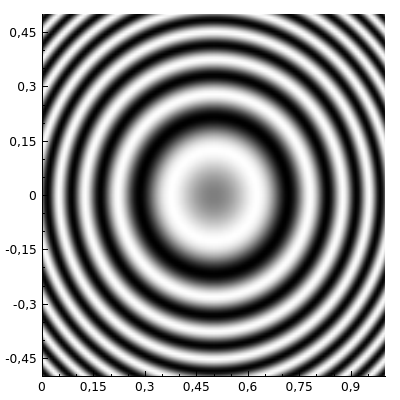} \\
			a) & b) & c)
		\end{tabular}
		\caption{
			\textbf{\ }\emph{a) Initial condition }$f\left( x\right) .$\emph{\
			b) Boundary conditions }$g\left( t\right) $ and $p\left( t\right) $\emph{.
			c) }$F\left( x,t\right) .$
			\label{fig:cond}
		}	
	\end{center}
\end{figure}

Graphs of functions $F,f,g,p$ are presented on Figure 1. Thus, solving the
forward problem (\ref{5.1})-(\ref{5.30}) for the input functions (\ref{5.4}%
)-(\ref{5.7}), we have computed the function $q_{comp}\left( t\right) ,$%
\begin{equation}
u_{x}\left( 1,t\right) =q_{comp}\left( t\right) ,t\in \left( -1/2,1/2\right)
.  \label{5.8}
\end{equation}%
We now formulate precisely the ill-posed Cauchy problem which we have solved
computationally.

\textbf{Ill-Posed Cauchy Problem 2}. \emph{Suppose that in (\ref{5.1})-(\ref%
{5.30}) functions }$f\left( x\right) $\emph{\ and }$g\left( t\right) $\emph{%
\ are unknown whereas the functions }$F\left( x,t\right) $\emph{\ }$p\left(
t\right) $\emph{, }$S\left( u\right) $\emph{\ and the constant }$a\geq 0$%
\emph{\ are known. Suppose that in the data simulation process functions }$%
F,S,f,g,p$\emph{\ are the same as in (\ref{5.4})-(\ref{5.7}). Determine the
function }$u\left( x,t\right) $\emph{\ in at least a subdomain of the time
cylinder }$Q_{1/2}^{\pm }$\emph{\ assuming that the function }$%
q_{comp}\left( t\right) $\emph{\ in (\ref{5.8}) is known. }

We now briefly describe how did we solve the forward problem (\ref{5.1})-(%
\ref{5.30}) numerically using FDM. Introduce the uniform mesh in the domain $%
Q_{1/2}^{\pm },$%
\begin{equation*}
\overline{M}=\left\{ \left( x_{i},t_{j}\right) :x_{i}=ih,t_{j}=-\frac{1}{2}%
+j\tau ,i\in \left[ 0,N\right) ,j\in \left[ 0,M\right) \right\} ,
\end{equation*}%
where $h=1/N$ and $\tau =1/M$ are grid step sizes in $x$ and $t$ directions
respectively. We have used $N=32,M=128.$ For generic functions $f^{\left(
1\right) }\left( x,t\right) ,f^{\left( 2\right) }\left( x\right) ,f^{\left(
3\right) }\left( t\right) $ denote $f_{ij}^{\left( 1\right) }=f^{\left(
1\right) }\left( x_{i},t_{j}\right) ,f_{i}^{\left( 2\right) }=f^{\left(
2\right) }\left( x_{i}\right) ,f_{j}^{\left( 3\right) }=f^{\left( 3\right)
}\left( t_{j}\right) .$ Let $\varphi _{ij}=aS\left( u_{ij}\right) +F_{ij}.$
We have solved the forward problem (\ref{5.1})-(\ref{5.30}) using the
implicit finite difference scheme,%
\begin{equation*}
\frac{u_{ij+1}-u_{ij}}{\tau }=\frac{1}{h^{2}}\left(
u_{i-1j+1}-2u_{ij+1}+u_{i+1j+1}\right) +\varphi _{ij},i\in \left[
1,N-1\right) ,j\in \left[ 0,M-1\right) ,
\end{equation*}%
\begin{equation*}
u_{i0}=f_{i},\text{ }u_{0j}=g_{j},u_{nj}=p_{j},i\in \left[ 0,N\right) ,j\in %
\left[ 0,M\right) ,
\end{equation*}

\subsection{Specifying the functional $J_{\protect\lambda ,\protect\beta }$}

\label{sec:5.2}

In the case of (\ref{5.1})-(\ref{5.8}) the operator $A$ becomes%
\begin{equation*}
K\left( u\right) =u_{t}-u_{xx}-aS\left( u\right) -F\left( x,t\right) .
\end{equation*}%
And the functional $J_{\lambda ,\beta }$ is%
\begin{equation}
J_{\lambda ,\beta }\left( u\right)
=\dint\limits_{-1/2}^{1/2}\dint\limits_{0}^{1}\left[ K\left( u\right) \right]
^{2}\varphi _{\lambda }^{2}dxdt+\beta \left\Vert u\right\Vert _{H^{2}\left(
Q_{1/2}^{\pm }\right) }^{2}.  \label{5.9}
\end{equation}%
We have dropped here the multiplier $e^{-2\lambda \left( \alpha
_{0}+\varepsilon \right) }$ which was present in the original version (\ref%
{2.13}). Indeed, we have used this multiplier above in order to allow the
parameter $\beta $ to be less than $1$. However, we have discovered in our
computations that the accuracy of results does not change much for $\beta $
varying in a large interval. The norm $\left\Vert u\right\Vert _{H^{2}\left(
Q_{1/2}^{\pm }\right) }$ is taken instead of $\left\Vert u\right\Vert
_{H^{3}\left( Q_{1/2}^{\pm }\right) }$ due to the convenience of
computations. Note that since we do not use too many grid points when
discretizing the functional $J_{\lambda ,\beta }\left( u\right) ,$ then
these two norms are basically equivalent in our computations, since all
norms in a finite dimensional space are equivalent.

\subsection{The discrete form of $J_{\protect\lambda ,\protect\beta }$}

\label{sec:5.3}

In our computations we represent derivatives in (\ref{5.9}) in the form of
finite differences with $N=32,M=128$ and minimize the resulting functional
with respect to values of the function $u$ at grid points. Discretizing
integrals, we obtain the following discrete form $\hat{J}$ of the functional
(\ref{5.9})%
\begin{equation}
\hat{J}(\hat{u})=\frac{1}{NM}\Big[\sum\limits_{i=1}^{N-2}\sum%
\limits_{j=0}^{M-2}K_{ij}^{2}\varphi _{\lambda ij}^{2}+\beta
\sum\limits_{i=1}^{N-2}\sum\limits_{j=1}^{M-2}Y_{ij}\Big],  \label{5.10}
\end{equation}%
where $\hat{u}=\{u_{00},u_{10},...,u_{kn},...,u_{N-1M-1}\}$ is the vector of
values of the function $u$ at grid points. Here\textbf{\ } 
\begin{equation*}
K_{ij}=\frac{u_{ij+1}-u_{ij}}{\tau }-\frac{u_{i-1j}-2u_{ij}+u_{i+1j}}{h^{2}}%
-aS\left( u_{ij}\right) -F_{ij},
\end{equation*}%
\begin{equation}
Y_{ij}=u_{ij}^{2}+\frac{\left( u_{ij+1}-u_{ij}\right) ^{2}}{\tau ^{2}}+\frac{%
\left( u_{i+1j}-u_{ij}\right) ^{2}}{h^{2}}  \label{5.11}
\end{equation}%
\begin{equation*}
+\frac{\left( u_{ij-1}-2u_{ij}+u_{ij+1}\right) ^{2}}{\tau ^{4}}+\frac{\left(
u_{i-1j}-2u_{ij}+u_{i+1j}\right) ^{2}}{h^{4}}.
\end{equation*}%
To apply the conjugate gradient method (GCM), it is convenient to use
explicit formulae for the derivatives $\partial \hat{J}(\hat{u})/\partial
u_{kn}.$ Using (\ref{5.10}), we obtain for indexes $1\leq k\leq N-2,1\leq
n\leq M-2$ 
\begin{equation}
\frac{\partial \hat{J}}{\partial u_{kn}}=\frac{2}{NM}\sum\limits_{i=1}^{N-2}%
\sum\limits_{j=0}^{M-2}\varphi _{ij}^{2}K_{ij}\frac{\partial K_{ij}}{%
\partial u_{kn}}+\frac{\beta }{NM}\sum\limits_{i=0}^{N-2}\sum%
\limits_{j=1}^{M-2}\frac{\partial Y_{ij}}{\partial u_{kn}}.  \label{5.12}
\end{equation}%
We calculate these derivatives only with respect to those parameters $u_{kn}$
which correspond to internal grid points, i.e. for above indices. We set 
\begin{equation}
\frac{\partial \hat{J}}{\partial u_{0j}}=0,\frac{\partial \hat{J}}{\partial
u_{i0}}=\frac{\partial \hat{J}}{\partial u_{iM-1}}=0.  \label{5.120}
\end{equation}%
Also, we set to zero partial derivatives of $\hat{J}$ with respect to $%
u_{N-1j}$ and $u_{N-2j}.$ This is because values of $u_{N-1j}$ and $u_{N-2j} 
$ are known, see (\ref{5.15}) and (\ref{5.16}). So, 
\begin{equation*}
\frac{\partial \hat{J}}{\partial u_{N-2j}}=\frac{\partial \hat{J}}{\partial
u_{N-1j}}=0
\end{equation*}

To simplify notations, we omit here and below the subscript $\lambda $ in$%
\varphi _{ij}^{2}.$ Using (\ref{5.11}), we obtain%
\begin{equation*}
\sum\limits_{i=1}^{N-2}\sum\limits_{j=0}^{M-2}\varphi _{ij}^{2}K_{ij}\frac{%
\partial K_{ij}}{\partial u_{kn}}=
\end{equation*}%
\begin{equation}
\frac{2}{\tau }\left( \varphi _{kn-1}^{2}K_{kn-1}-\varphi
_{kn}^{2}K_{kn}\right) -\frac{2}{h^{2}}\Big(\varphi
_{k-1n}^{2}K_{k-1n}-2\varphi _{kn}^{2}K_{kn}+\varphi _{k+1n}^{2}K_{k+1n}\Big)
\label{5.13}
\end{equation}%
\begin{equation*}
-2\varphi _{kn}^{2}K_{kn}^{2}S^{\prime }(u_{kn}),
\end{equation*}%
where $S^{\prime }(u_{kn})$ is determined by the function $S(u)$ and can be
calculated analytically. Next, 
\begin{equation*}
\sum\limits_{i=0}^{N-2}\sum\limits_{j=1}^{M-2}\frac{\partial Y_{ij}}{%
\partial u_{kn}}=
\end{equation*}%
\begin{eqnarray}
&&2u_{kn}+\frac{2}{\tau }\Big(Ut_{kn+1}-Ut_{kn}\Big)+\frac{2}{h}\Big(%
Ux_{k+1n}-Ux_{kn}\Big)+  \label{5.14} \\
&&+\frac{2}{\tau ^{2}}\Big(Utt_{kn+1}-2Utt_{kn}+Utt_{kn-1}\Big)++\frac{2}{%
h^{2}}\Big(Uxx_{k+1n}-2Uxx_{kn}+Uxx_{k-1n}\Big),  \notag
\end{eqnarray}%
where 
\begin{equation*}
Ut_{kn}=\left\{ 
\begin{array}{c}
\frac{1}{\tau }(u_{kn+1}-u_{kn})\quad \text{if}\quad n\in \lbrack 0,M-1) \\ 
0\quad \text{if}\quad n=M-1;%
\end{array}%
\right.
\end{equation*}%
\begin{equation*}
Ux_{kn}=\left\{ 
\begin{array}{c}
\frac{1}{h}(u_{k+1n}-u_{kn})\quad \text{if}\quad k\in \lbrack 0,N-1) \\ 
0\quad \text{if}\quad k=N-1;%
\end{array}%
\right.
\end{equation*}%
\begin{equation*}
Utt_{kn}=\left\{ 
\begin{array}{c}
\frac{1}{\tau ^{2}}(u_{kn+1}-2u_{kn}+u_{kn-1})\quad \text{if}\quad n\in
(0,M-1) \\ 
0\quad \text{if}\quad n=0\quad \text{and}n=M-1;%
\end{array}%
\right.
\end{equation*}%
\begin{equation*}
Uxx_{kn}=\left\{ 
\begin{array}{c}
\frac{1}{h^{2}}(u_{k+1n}-2u_{kn}+u_{k-1n})\quad \text{if}\quad k\in (0,N-1)
\\ 
0\quad \text{if}\quad k=0\quad \text{and}k=N-1;%
\end{array}%
\right.
\end{equation*}

In (\ref{5.13}) and (\ref{5.14}) we use boundary conditions (\ref{5.30}) and
(\ref{5.8}) at $x=1$ as 
\begin{equation}
u_{N-1j}=p_{j},u_{N-2j}=p_{j}-hq_{comp,j}.  \label{5.15}
\end{equation}

\subsection{Some notes about noisy data and the conjugate gradient method}

\label{sec:5.4}

In all our numerical experiments $\beta =0.00063.$ As we have stated in
subsection 5.2, we have observed in our computations that this parameter
does not influence much our results. All results below are obtained for
noisy data with 5\% level of noise. Here is how we have introduced this
noise. Let $\sigma \in \left[ -1,1\right] $ be the random variable
representing the white noise. Let $p^{\left( m\right) }=\max_{j}\left\vert
p_{j}\right\vert $ and $q^{\left( m\right) }=\max_{j}\left\vert
q_{comp,j}\right\vert .$ Then by (\ref{5.15}) the noisy data, which we have
used, were%
\begin{equation}
\widetilde{u}_{N-1j}=p_{j}+0.05p^{\left( m\right) }\sigma _{j},\widetilde{u}%
_{N-2j}=p_{j}-h\left( q_{comp,j}+0.05q^{\left( m\right) }\sigma _{j}\right) .
\label{5.16}
\end{equation}

In all our numerical tests we have used in (\ref{5.10}) $M=32,N=128$. Even
though these numbers are the same as in the solution of the forward problem,
the \textquotedblleft inverse crime" was not committed since we have used
noisy data and since we have used the minimization of the functional (\ref%
{5.10}) rather than solving a forward problem again. To minimize the
functional, we have used the unconstrained CGM. We arrange this method in
such a way that boundary conditions (\ref{5.16}) are kept to be satisfied on
all iterations. So, we minimize the functional (\ref{5.10}) with respect to
numbers $\left\{ u_{ij}\right\} _{\left( i,j\right) =\left( 2,2\right)
}^{\left( N-3,M-3\right) }.$ However, numbers $u_{N-1j},u_{N-2j}$ are kept
as (\ref{5.16}). As the starting vector $\left\{ u_{ij}^{0}\right\} $ we take%
\begin{equation}
u_{ij}^{0}=\left\{ 
\begin{array}{c}
0\text{ if }i\in \left[ 0,N-3\right] , \\ 
\widetilde{u}_{N-2j}\text{ if }i=N-2, \\ 
\widetilde{u}_{N-1j}\text{ if }i=N-1.%
\end{array}%
\right.  \label{5.17}
\end{equation}

Normally, for a quadratic functional this method reaches its minimum after $%
M\cdot N$ gradient steps with the automatic step choice. However, our
computational experience tells us that we can obtain a better accuracy if
using a small constant step in the GCM and a constant number of iterations.
Thus, we have used the step size $\gamma =10^{-8}$ and 10,000 iterations of
the GCM. It took 0.5 minutes of CPU Intel Core i7 to do these iterations.

\section{Numerical Results}

\label{sec:6}

Let $u\left( x,t\right) $ be the numerical solution of the forward problem (%
\ref{5.1})-(\ref{5.3}). Let $u_{\lambda \beta }\left( x,t\right) $ be the
minimizer of the functional (\ref{5.10}) which we have found via GCM. Of
course, $u\left( x,t\right) $ and $u_{\lambda \beta }\left( x,t\right) $
here are discrete functions defined on the above grid and norms used below
are discrete norms. For each $x$ from this grid we define the
\textquotedblleft line error" $E\left( x\right) $ as%
\begin{equation}
E\left( x\right) =\frac{\left\Vert u_{\lambda \beta }\left( x,t\right)
-u\left( x,t\right) \right\Vert _{L_{2}\left( -1/2,1/2\right) }}{\left\Vert
u\left( x,t\right) \right\Vert _{L_{2}\left( -1/2,1/2\right) }}.  \label{6.1}
\end{equation}%
We evaluate how the line error changes with the change of $x$. Since our
lateral data are given at $x=1$, it is anticipated that the function $%
E\left( x\right) $ should be decreasing.

We have tested three values of the parameter $\lambda :\lambda =0,3,4.$ We
have found that $\lambda =4$ is the best choice, at least for those problems
which we have studied. Also, we have tested two values of the parameter $a$
in (\ref{5.1}): $a=0$ and $a=10$. The case $a=0$ corresponds to the linear
problem and $a=10$ indicates the nonlinearity.

\subsection{Graphs of line errors}

\label{sec:6.1}

Graphs of the line error are presented on Figure 2. Figure 2a corresponds to
the linear case with $a=0$. Figures 2b and 2c with $a=10$ display the line
errors for two above functions $S\left( u\right) .$ Figure 2b is for $%
S\left( u\right) =\sin ^{2}\left( u\right) $ and Figure 2c is for $S\left(
u\right) =\exp \left( 0.4u\right) .$ One can observe that for both values $%
\lambda =3$ and $\lambda =4$ all three cases have an acceptable error up to $%
x=0.45.$ In other words, the function $u\left( x,t\right) $ is reconstructed
rather accurately on more than half of the interval $\left[ 0,1\right] $:
for $x\in \left[ 0.45,1\right] .$ Thus, in all cases the presence of the CWF
significantly improves the accuracy of the solution. Furthermore, the
presence of the CWF improves the accuracy even in the linear case.

The case $a=0,\lambda =0$ corresponds to the Quasi-Reversibility Method,
which was first introduced by Lattes and Lions \cite{LL}. This method works
only for linear PDEs. The convergence rate of this method can be established
via Carleman estimates, see \cite{B1,B2,D1,D2,KT} and the recent survey \cite%
{APNUM}.

\begin{figure}[hb!]
	\begin{center}
		\begin{tabular}{c}
			\includegraphics[scale=0.55,clip=]{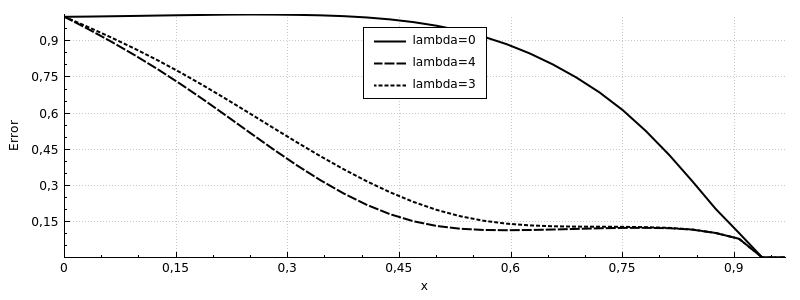} \\
			a)\\
			\includegraphics[scale=0.55,clip=]{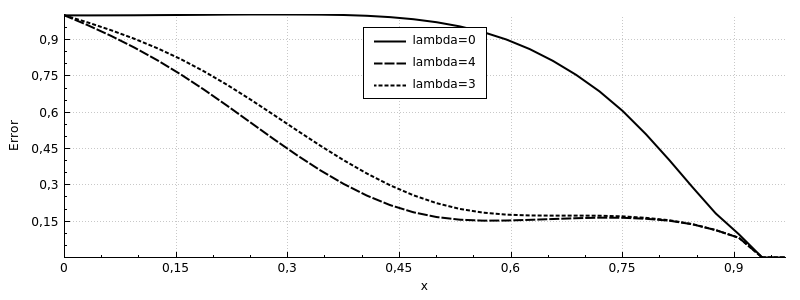} \\
			b)\\
			\includegraphics[scale=0.55,clip=]{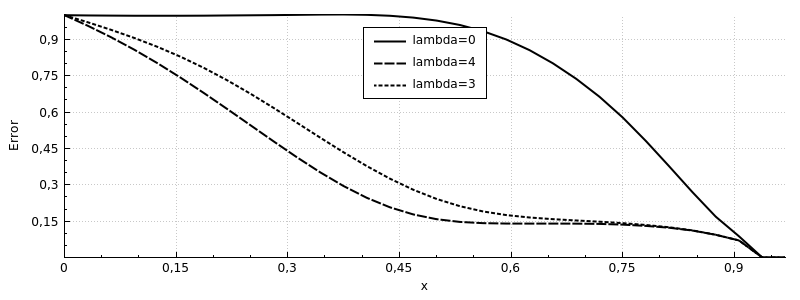} \\
			c)
		\end{tabular}
		\caption{ 
			\emph{Line errors. a) The linear problem with }$a=0$\emph{\ for }$%
			\lambda =0,3,4.$\emph{\ The case }$a=0,\lambda =0$\emph{\ corresponds to the
			Quasi-Reversibility Method. b) }$a=10,\lambda =0,3,4$\emph{\ and }$S\left(
			u\right) =\sin ^{2}\left( u\right) .$\emph{\ c) }$a=10,\lambda =0,3,4$\emph{%
			\ and }$S\left( u\right) =\exp \left( 0.4u\right) .$\emph{\ Thus, in all
			cases the presence of the CWF in the functional (\ref{5.10}) significantly
			improves the accuracy of the solution. Furthermore, the presence of the CWF
			improves the accuracy even in the linear case. In all cases a rather
			accurate reconstruction is obtained for }$x\in \left[ 0.45,1\right] ,$\emph{%
			\ i.e. on more than half of the interval }$\left[ 0,1\right] .$
			\label{fig:line_errors}
		}
	\end{center}
\end{figure}

\subsection{Graphs of functions $u_{\protect\lambda \protect\beta }\left(
0.6,t\right) $}

\label{sec:6.2}

As one can see on Figures 2 a)-c), the line error at $x=0.6$ is about 10\%
for $\lambda =0,4$ for all three cases. Thus, we have decided to superimpose
graphs of functions $u_{\lambda \beta }\left( 0.6,t\right) $ with graphs of
functions $u\left( 0.6,t\right) .$ One can see on Figures 3a)-c) that graphs
of functions $u_{0\beta }\left( 0.6,t\right) $ are rather far from the graph
of functions $u\left( 0.6,t\right) .$ On the other hand, the presence of the
CWF in the functional (\ref{5.10}) makes graphs of functions $u_{4\beta
}\left( 0.6,t\right) $ to be quite close to the graph functions $u\left(
0.6,t\right) .$ This is true even for the linear case of Figure 3a). On the
other hand, functions $u_{0\beta }\left( 0.6,t\right) $ and $u_{4\beta
}\left( 0.6,t\right) $ drop to zero as $t\rightarrow -1/2^{+}.$ Also, the
accuracy at $t\approx 1/2$ is not good on Figures 3a),b). We explain this by
condition (\ref{5.120}) which we have imposed. In addition, in the accuracy
estimates (\ref{2.22}), (\ref{2.23}) $G_{\alpha _{0}+2\varepsilon }\cap
\left\{ t=\pm 1/2\right\} =\varnothing .$

\begin{figure}[hb!]
	\begin{center}
		\begin{tabular}{c}
			\includegraphics[scale=0.53,clip=]{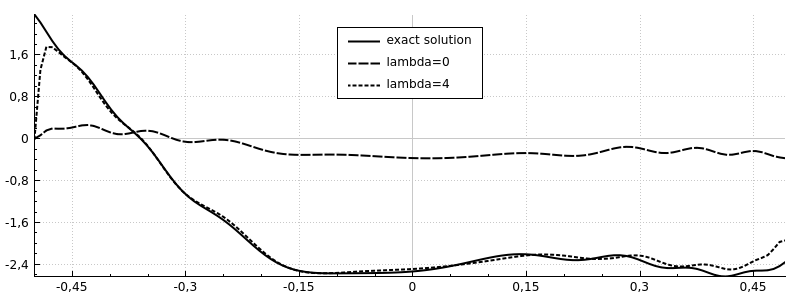} \\
			a)\\
			\includegraphics[scale=0.53,clip=]{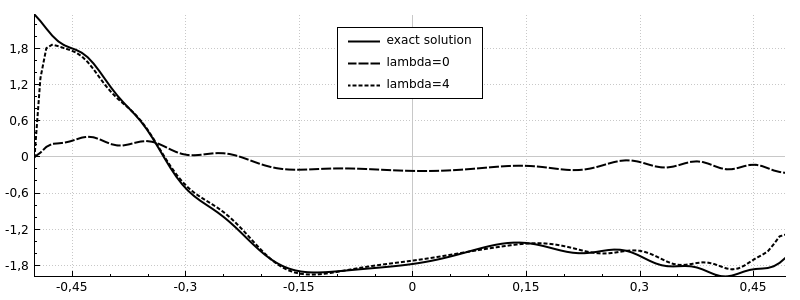} \\
			b)\\
			\includegraphics[scale=0.53,clip=]{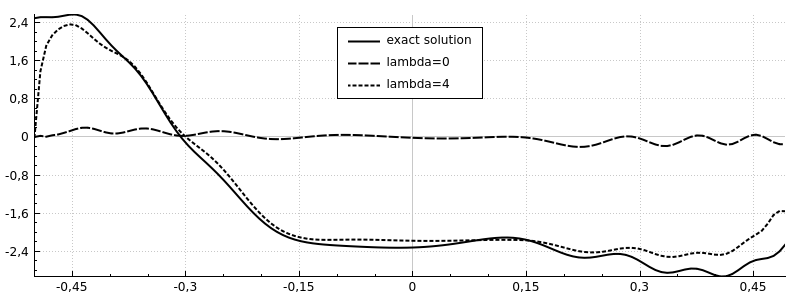} \\
			c)
		\end{tabular}
		\caption{
			\emph{Superimposed graphs of functions }$u_{0\beta }\left(
			0.6,t\right) ,u_{4\beta }\left( 0.6,t\right) $\emph{\ and }$u\left(
			0.6,t\right) .$\emph{\ a) The linear case, }$a=0$\emph{. b) }$a=10,S\left(
			u\right) =\sin ^{2}\left( u\right) .$\emph{\ c) }$a=10,S\left( u\right)
			=\exp \left( 0.4u\right) .$\emph{\ Observe that the presence of the CWF with 
			}$\lambda =4$\emph{\ quite essentially improves the accuracy of the solution
			in all three cases, including even the linear case.}
			\label{fig:slices}
		}	
	\end{center}
\end{figure}

\subsection{The influence of the initial condition}

\label{sec:6.3}

To see how the knowledge of the initial condition $u\left( x,-1/2\right)
=f\left( x\right) $ affects the accuracy of our results, we have tested the
case when the function $f\left( x\right) $ in (\ref{5.6}) is known.\emph{\ }%
Now\emph{\ }we arrange the GCM in such a way that both boundary conditions (%
\ref{5.16}) and the initial condition $u_{i0}=f_{i}$ are kept be satisfied
on all iterations. Similarly with (\ref{5.17}) the first guess $\left\{
u_{ij}^{0}\right\} $ is taken as

\begin{equation*}
u_{ij}^{0}=\left\{ 
\begin{array}{c}
0\text{ if }i\in \left[ 0,N-3\right] ,j\neq 0, \\ 
\widetilde{u}_{N-2j}\text{ if }i=N-2,j\neq 0, \\ 
\widetilde{u}_{N-1j}\text{ if }i=N-1,j\neq 0, \\ 
f_{i}\text{ if }j=0.%
\end{array}%
\right.
\end{equation*}

\begin{figure}[hb!]
	\begin{center}
		\begin{tabular}{c}
			\includegraphics[scale=0.51,clip=]{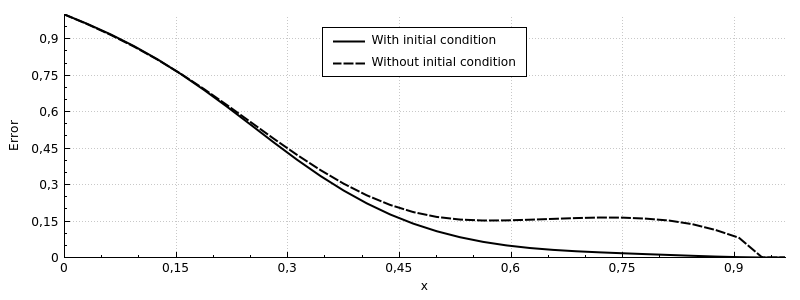} \\
			a)\\
			\includegraphics[scale=0.51,clip=]{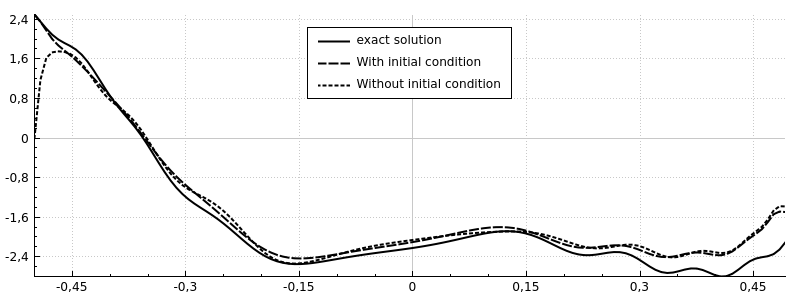} \\
			b)\\
			\includegraphics[scale=0.51,clip=]{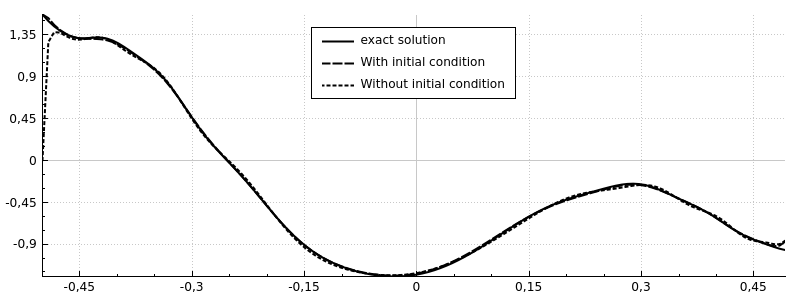} \\
			c)
		\end{tabular}
		\caption{
			\emph{The influence of the knowledge of the initial condition. The
			case }$a=10,S\left( u\right) =exp(0.4u)\left( u\right) .$\emph{\ a) Line
			errors. b) Graphs of functions }$u_{4\beta }\left( 0.6,t\right) $\emph{\
			with and without knowledge of the initial condition, superimposed with the
			graph of the function }$u\left( 0.6,t\right) .$\emph{\ c) The same as in b)
			but for functions }$u_{4\beta }\left( 0.8,t\right) ,u\left( 0.8,t\right) .$%
			\emph{\ One can see that the knowledge of the initial condition does not
			provide an essential impact in the accuracy of the solution.}
			\label{fig:initial_cond_influence}
		}	
	\end{center}
\end{figure}

Here we consider the case $a=10,S\left( u\right) =\sin ^{2}\left( u\right) .$
Figure 4a) displays the line errors with and without knowledge of the
function $f\left( x\right) $ in (\ref{5.6}). One can see that for $x\in %
\left[ 0.45,1\right] $ the error for the case when $f\left( x\right) $ is
known is less than for the case when $f\left( x\right) $ is unknown. Figure
4b) displays graphs of functions $u_{4\beta }\left( 0.6,t\right) $ for the
cases of known and unknown initial condition. They are superimposed with the
graph of the function $u\left( 0.6,t\right) .$ One can observe that these
three graphs are only slightly different from each other on the major part
of the time interval. \ Figure 4c) displays graphs of functions $u_{4\beta
}\left( 0.8,t\right) $ for the cases of known and unknown initial condition.
They are superimposed with the graph of the function $u\left( 0.8,t\right) .$
One can see that these graphs almost coincide, except of $t\approx -1/2.$
Thus, the knowledge of the initial condition does not provide an essential
impact in the accuracy of the solution.

\begin{center}
\textbf{Acknowledgments}
\end{center}

The work of the first author was supported by US Army Research Laboratory
and US Army Research Office grant W911NF-15-1-0233 and by the Office of
Naval Research grant N00014-15-1-2330.

\end{document}